\documentclass[amssymb,aps,prb,floats,twocolumn,superscriptaddress,showpacs]{revtex4-1}
\usepackage{graphicx}
\usepackage{graphics}
\usepackage{bm}
\usepackage{tabularx}
\usepackage{epstopdf}
\usepackage{array}
\usepackage{float}
\usepackage{units}

\begin{document}

\title{Magnetoresistance in Fe$_{1-x}$Ga$_x$ thin films presenting striped magnetic pattern: the role of closure domains and domain walls}

\author{B. Pianciola}
\affiliation{Instituto de Nanociencia y Nanotecnolog\'ia, CNEA-CONICET, Centro At\'omico Bariloche, (R8402AGP) San Carlos de Bariloche, Argentina}

\author{S. Flewett}
\affiliation{Instituto de F\'isica, Pontificia Universidad Cat\'olica de Valpara\'iso, Avenida Universidad 330, Valpara\'iso, Chile}

\author{E. De Biasi}
\affiliation{Instituto de Nanociencia y Nanotecnolog\'ia, CNEA-CONICET, Centro At\'omico Bariloche, (R8402AGP) San Carlos de Bariloche, Argentina}
\affiliation{Universidad Nacional de Cuyo, Instituto Balseiro, Centro At\'omico Bariloche, (R8402AGP) San Carlos de Bariloche, Argentina.}

\author{C. Hepburn}
\affiliation{Sorbonne Universit\'e, CNRS, Institut des NanoSciences de Paris, INSP, UMR7588, F-75005 Paris, France}

\author{L. Lounis}
\affiliation{Sorbonne Universit\'e, CNRS, Institut des NanoSciences de Paris, INSP, UMR7588, F-75005 Paris, France}

\author{M. V\'asquez-Mansilla}
\affiliation{Instituto de Nanociencia y Nanotecnolog\'ia, CNEA-CONICET, Centro At\'omico Bariloche, (R8402AGP) San Carlos de Bariloche, Argentina}

\author{M. Granada}
\affiliation{Instituto de Nanociencia y Nanotecnolog\'ia, CNEA-CONICET, Centro At\'omico Bariloche, (R8402AGP) San Carlos de Bariloche, Argentina}

\author{M. Barturen}
\affiliation{Instituto de Tecnolog\'ia, Universidad Argentina de la Empresa, 
Lima 775, (C1073AAO) Ciudad Aut\'onoma de Buenos Aires, Argentina}

\author{M. Eddrief}
\affiliation{Sorbonne Universit\'e, CNRS, Institut des NanoSciences de Paris, INSP, UMR7588, F-75005 Paris, France}
\affiliation{LIFAN, Laboratorio Internacional Franco-Argentino en Nanociencias}

\author{M. Sacchi}
\affiliation{Sorbonne Universit\'e, CNRS, Institut des NanoSciences de Paris, INSP, UMR7588, F-75005 Paris, France}
\affiliation{LIFAN, Laboratorio Internacional Franco-Argentino en Nanociencias}
\affiliation{Synchrotron SOLEIL, L'Orme des Merisiers, Saint-Aubin BP 48, 91192 Gif-sur-Yvette Cedex, France}

\author{M. Marangolo}
\affiliation{Sorbonne Universit\'e, CNRS, Institut des NanoSciences de Paris, INSP, UMR7588, F-75005 Paris, France}
\affiliation{LIFAN, Laboratorio Internacional Franco-Argentino en Nanociencias}

\author{J. Milano}
\affiliation{Instituto de Nanociencia y Nanotecnolog\'ia, CNEA-CONICET, Centro At\'omico Bariloche, (R8402AGP) San Carlos de Bariloche, Argentina}
\affiliation{Universidad Nacional de Cuyo, Instituto Balseiro, Centro At\'omico Bariloche, (R8402AGP) San Carlos de Bariloche, Argentina.}
\affiliation{LIFAN, Laboratorio Internacional Franco-Argentino en Nanociencias}

\date{\today}

\begin{abstract}

In this work we show the existence of closure domains in Fe$_{1-x}$Ga$_x$ thin films featuring a striped magnetic pattern and study the effect of the magnetic domain arrangement on the magnetotransport properties. By means of X-ray resonant magnetic scattering, we experimentally demonstrate the presence of such closure domains and estimate their sizes and relative contribution to surface magnetization. Magnetotransport experiments show that the behavior of the magnetoresistance depends on the measurement geometry as well as on the temperature. When the electric current flows perpendicular to the stripe direction, the resistivity decreases when a magnetic field is applied along the stripe direction (negative magnetoresistance) in all the studied temperature range, and the calculations indicate that the main source is the anisotropic magnetoresistance. In the case of current flowing parallel to the stripe domains, the magnetoresistance changes sign, being positive at room temperature and negative at 100~K. To explain this behavior, the contribution to magnetoresistance from the domain walls must be considered besides the the anisotropic one.

\end{abstract}
\maketitle

\section{Introduction}

Ferromagnetic thin films characterized by weak stripe domain patterns feature some peculiar properties that open perspectives for their use in magnonic devices. Recently, C. Liu et al.\cite{Liu2019} have shown that in these films it is possible to control the magnetization direction by electrical currents, injecting current densities one order of magnitude lower than what is typically needed when relying on the spin torque effect. They observed also that the attenuation of a magnon current is highly dependent on its direction with respect to the stripe orientation.

Some years ago the presence of striped magnetic pattern was reported in Fe$_{1-x}$Ga$_x$ thin films\cite{marianaAPL,mariana_ani} and was ascribed to the presence of a perpendicular-to-the-film magnetic anisotropy (PMA). PMA induces a magnetic easy axis along the surface normal and leads to the appearence of a magnetic structure within the film that is more complex than the usual in-plane configuration imposed by magnetostatics\cite{saito}. The $Q$ parameter (defined as $Q$~=~$\nicefrac{2 K_{\rm PMA}}{\mu_0 M^2}$, where $K_{\rm PMA}$ is the strength of the PMA and $\nicefrac{\mu_0 M^2}{2}$ is the demagnetizing energy, $E_{\rm dem}$, for a magnetic thin film) helps us to measure how far the system is from a fully in-plane magnetic configuration, $Q < 0$. If $Q > 1$, $K_{\rm PMA}$ overcomes $E_{\rm dem}$, so the magnetization points perpendicular to the thin film. However, if $0< Q < 1$, an intermediate state exists where PMA competes with the magnetostatic energy. In this case, the film presents self organized stripe-shaped magnetic domains with a complex magnetic structure\cite{schaffer}. This periodically modulated arrangement, which can be controlled via a simple magnetic procedure\cite{hagostripes}, presents domains with the magnetization vectors pointing along the three spatial directions within the sample. A fraction of the magnetic moments remains along the direction of the most recent saturating field ($w$ domains), another fraction points perpendicular to the film plane ($s$ type) and, finally, closure domains ($c$) appear to reduce the stray magnetic field. The localization of each kind of magnetic domain is sketched in Fig.~\ref{fig:1-PreCharac}($a$). In spite of being predicted analytically\cite{murayama} and also obtained via micromagnetic calculations\cite{MarianaItalianos}, a quantitative  experimental procedure to study the complete domain structure to nanometer resolution has still not been fully developed. There has been much work done to characterize the domain structure of magnetic thin films in both 2- and 3-dimensions\cite{Donnelly2017,KORTRIGHT2013178} and references contained within, soft X-rays being frequently the probe of choice.

The interest in the magnetotransport properties of striped systems started in the '90 with the goal of studying giant magnetoresistance (GMR)\cite{Gregg1996}.
However, the GMR ratio is negligible because of the domain walls that separate the magnetic regions of those samples were not sharp enough to provoke a detectable spin dependent scattering. Nevertheless, it was the starting point for studying the different contributions to the magnetoresistance (MR) that are present in this type of systems, such as anisotropic magnetoresistance (AMR), Lorentz magnetoresistance (LMR) and domain wall magnetoresistance (DWMR)\cite{Marrows2005,kent2001domain}, being this last contribution to MR still a source of debate. Experimentally, DWMR does not always present the same behavior among the materials where it has been studied, specifically, it can present both positive or negative contributions to the resistivity\cite{Ruediger1998,Ruediger1999,Klein2000,PRB-Mara,Reeve2019}. Theoretically, semiclassical and quantum models were proposed in order to explain its origin using very different approaches, resulting in dissimilar predicted behaviors\cite{Viret1996,Levy1997,Tatara1997,vanGorkom1999,Gopar2004,Reeve2019}.

In Ref.~\onlinecite{PRB-Mara}, some of the authors have studied the magnetotransport properties of Fe$_{1-x}$Ga$_x$, in order to evaluate the behavior of the MR as a function of temperature in different measurement geometries [as depicted in Fig.~\ref{fig:AMR}($a$)]. We found that at room temperature the sign of the MR depends on the measurement geometry: for the case of electric current flowing perpendicular to stripe direction, the MR (CPW-MR) is negative while, when the current flows along the stripe direction, the MR (CIW-MR) is positive. Moreover, CIW-MR changes its sign at a temperature lower than 300~K.

In this work, we experimentally observe the existence and estimate the size of clousure domains in Fe$_{1-x}$Ga$_x$ thin films via X-ray resonant magnetic scattering (XRMS). We also study the electrical transport in the ohmic regime as a function of an applied field when stripes are present. Finally, by considering the magnetic configuration obtained by means of micromagnetic simulations, we correlate the existance of closure domains with the AMR behavior. From the analysis of the experimental data and simulations, we observed that the role played by the DWMR depends on the relative direction of the electric current with respect to stripes.

\section{Experimental details and preliminar characterization}

\begin{figure}
 \centering\includegraphics[width=1.0\columnwidth]{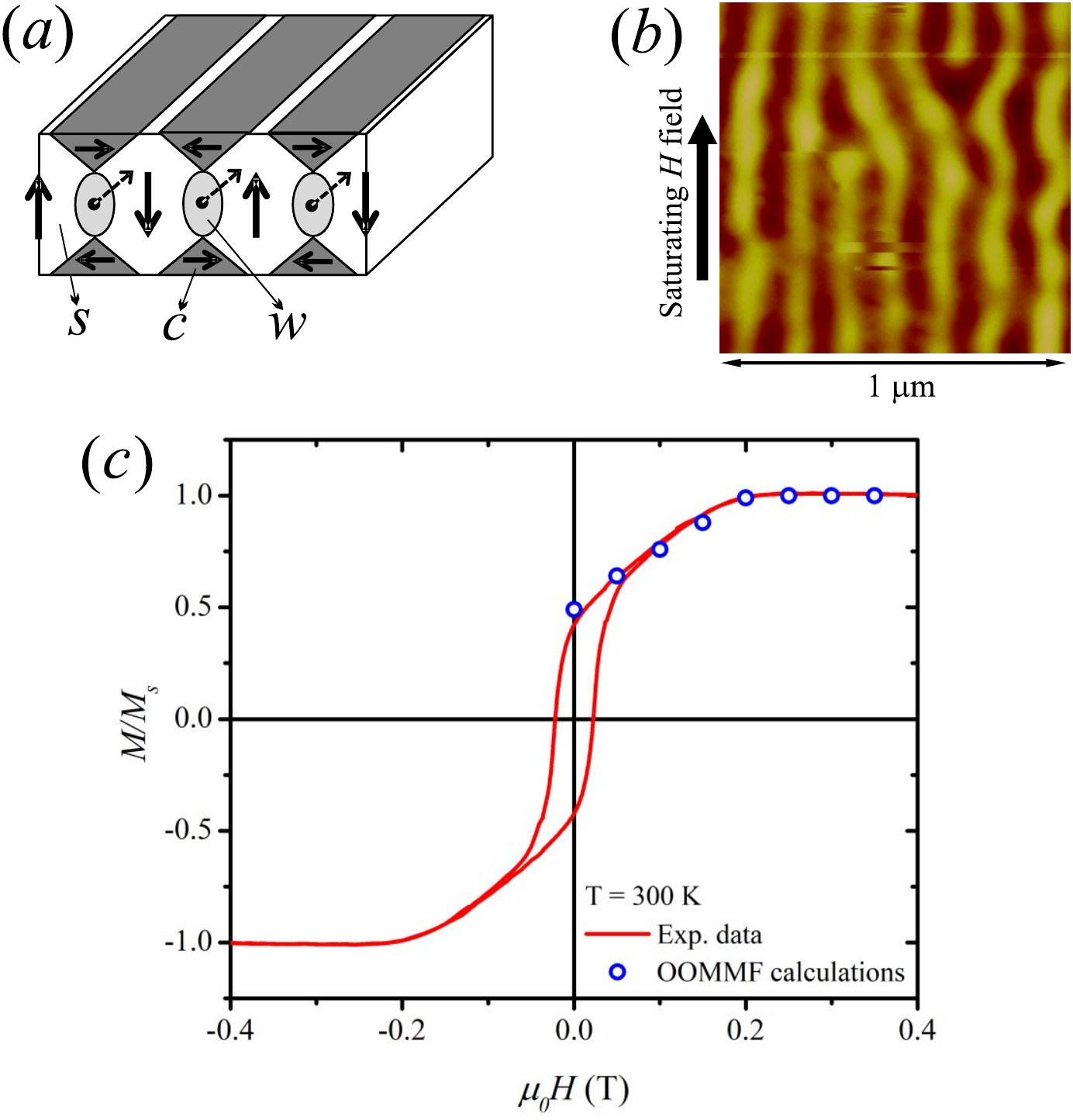}
\caption{\label{fig:1-PreCharac} ($a$) Schematic view of domains that compose the stripe pattern. ($b$) MFM image of the stray field generated by the stripe pattern. ($c$) $M$ vs. $H$ loop, where the characteristic linear behavior of the stripe domains at lower fields is observed in the experimental data (red line) as well as in the OOMMF calculations (blue circles).}
\end{figure}

Epitaxial Fe$_{1-x}$Ga$_x$ samples were grown by Molecular Beam Epitaxy on c(2x2) Zn-terminated ZnSe epilayers onto GaAs(001) 
substrates\cite{eddriefPRB2006,eddriefAPL2002}. After growth, the films were covered by a protective 3-nm gold capping layer. Details of the growth are given in Ref.~\onlinecite{Eddrief-PRB}. We fabricated 84-nm thick samples with a Ga concentration of $x$~=~0.20. Such concentration was determined by means of X-ray photoelectron spectroscopy (XPS) and confirmed by Rutherford backscattering (RBS) and energy dispersive X-ray spectrometry (EDX).
In order to determine the existence and estimate the size of closure domains in the Fe$_{1-x}$Ga$_x$ thin films, we performed XRMS experiments at the Circular Polarization beamline of the ELETTRA synchrotron, using the IRMA scattering chamber\cite{Sacchi2003}. The photon energy was set to 707 eV (Fe-2$p$ resonance) and we used circular polarization of both helicities produced using a helical electromagnetic wiggler source. The scattered intensity was measured using a two-dimensional (2D) detector formed by microchannel plates and a resistive anode\cite{BarturenEPJB}.

For the study of the magnetotransport properties, the electrical resistance was measured in a standard four-probe configuration, with collinear contacts along the [110] Fe$_{0.8}$Ga$_{0.2}$ crystalline direction. The voltage contacts were separated by 1.5~mm, so the effective size of the sample was much larger than the stripe period. The measurements were performed with a maximum DC electric current of 10 mA, which gives a current density of $\sim 0.1$ GA/m$^2$, much lower than the current densities needed to induce domain wall displacement (e.g., some TA/m$^2$ for Ni$_{81}$Fe$_{19}$ \cite{Torrejon}). Thus, we should not expect the electric current to affect the magnetic configuration in any way. The field dependence of the resistivity was measured with the magnetic field applied perpendicular and parallel to the electric current. Additional magnetization measurements were performed in a superconducting quantum interference device (SQUID) and a vibrating sample magnetometer (VSM). In Fig.~\ref{fig:1-PreCharac}($b$) and ($c$) we show preliminary magnetic characterization of the measured samples. Fig.~\ref{fig:1-PreCharac}($b$) shows a magnetic force microscopy (MFM) image. From this picture we can determine a spatial period of the stripes, $\lambda_S \sim 150$~nm. In Fig.~\ref{fig:1-PreCharac}($c$), we display the $M$~{\it vs}.~$H$ loop obtained by VSM at room temperature. We can observe the typical linear behavior at low field, which is a fingerprint of the presence of stripe domains. Also, we performed micromagnetic simulations via the OOMMF package\cite{OOMMF} in order to determine the magnetic structure of the studied systems.
\begin{figure}
 \centering\includegraphics[width=1.0\columnwidth]{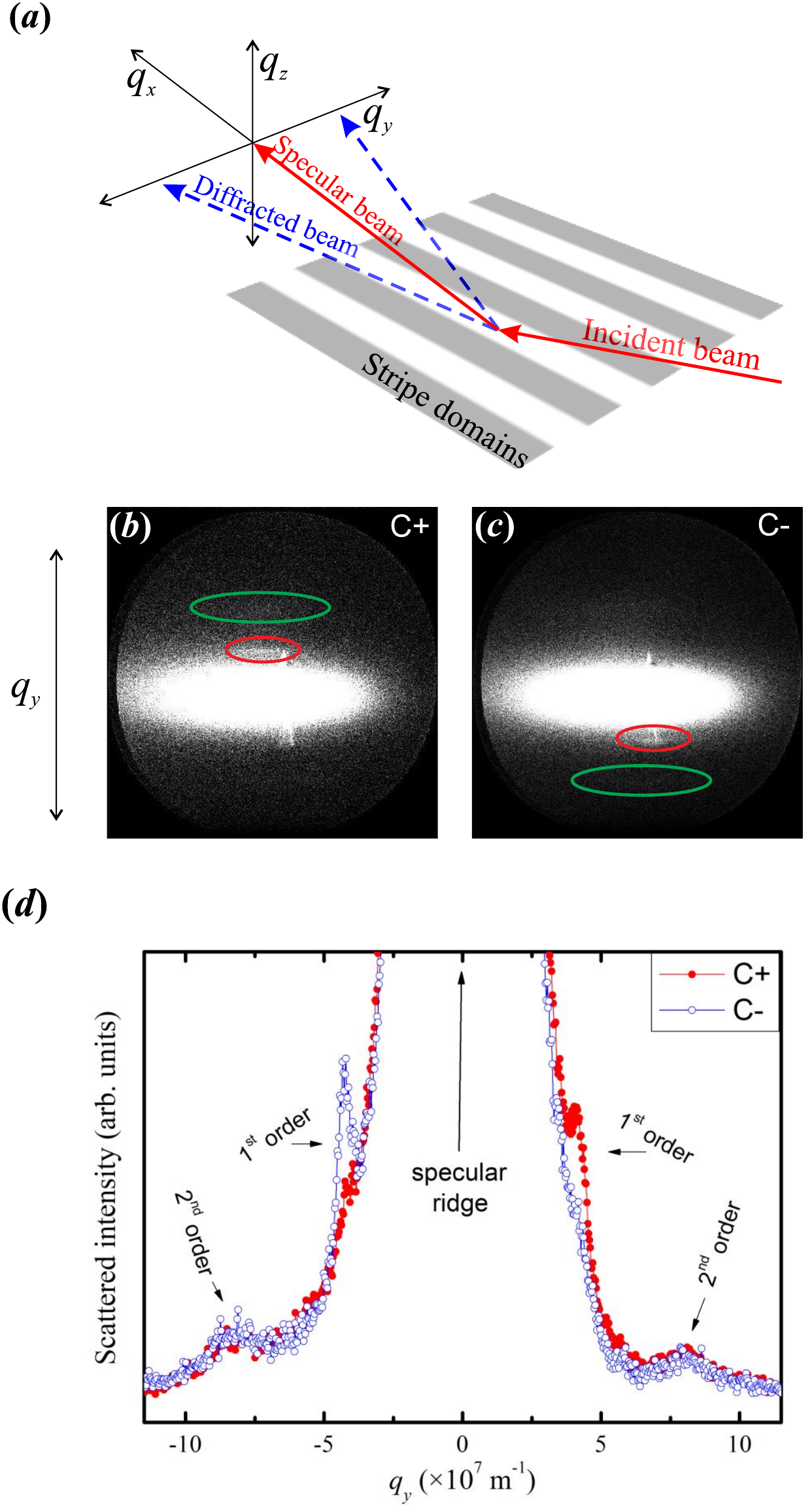}
\caption{\label{fig:1-rocking} ($a$) Measurement geometry. ($b$) and ($c$) images of the spots corresponding to the specular and diffracted peaks for C+ and C- respectively. ($d$) Peaks transverse section.}
\end{figure}
The saturation field ($H_s$) and the linear behavior at low field of the hysteresis loop were adjusted [as shown in Fig.~\ref{fig:1-PreCharac}($c$)] using the following parameters: $M_s$~=~1.4$\times$10$^5$~J/m$^3$, $K_{\rm PMA}$~=~3.5$\times$10$^5$~J/m$^3$,  and $A$~=~2.5$\times$10$^{-11}$~J/m. $M_s$ and $K_{\rm PMA}$ are in agreement with those experimentally obtained\cite{marianaAPL,mariana_ani}, while the adjusted $A$ value is larger than previously reported\cite{MarianaItalianos}. Using these parameters, we obtained a stripe pattern with a period of $\sim$140~nm. In Fig.~\ref{fig:1-PreCharac}, we show that $H_s$ and the remnant magnatization, $M_r$, obtained via OOMMF (blue circles) present a very good agreement with the experimental results. From the calculated magnetic structure we performed simulations in order to complement the experimental results obtained by XRMS and MR measurements. For the calculations we considered a sample 84 nm thick and with surface area 570$\times$570~nm$^2$, using periodic boundary conditions to avoid size effects along the sample plane. The sample was discretized into cubic cells of 3.5 nm, which is smaller than the magnetic exchange length\cite{Camara}. 

\section{Results and Discussion}
\subsection{Magnetic structure}

In Figs.~\ref{fig:1-rocking} ($a$), ($b$) and ($c$), we show the results obtained by means of XRMS experiments at the Fe-$L_3$ edge when the stripes are aligned within the scattering plane, using a circularly polarized incident beam with either positive (C+) or negative (C-) helicity. At the center of both pictures a large specular reflection peak is present\cite{expSR}. Marked in red one can observe the first order diffracted peaks. Note the very high degree of asymmetry observed between the opposite helicities, however given the partial swamping of the magnetic peaks by the dominant specular peak, the exact degree of this asymmetry is difficult to determine from our data. As previously demonstrated by D\"urr et al.\cite{durr}, this asymmetry is due to the presence of a mixed out of plane/in-plane magnetization at the sample surface (When XRMS is performed in reflection geometry, it probes the magnetic structure of a few nanometers from the surface). Furthermore, we can observe the appearence of second order peaks (marked in green) whose relative intensity appears not to depend on the beam polarization, although according to analysis  they are of magnetic origin and likely due to a partial protrusion of the Bloch wall to the surface layer probed by the X-rays.  
Fig.~\ref{fig:1-rocking}($d$) compares line profiles along $q_y$ drawn across the magnetic peaks of the 2D images in ($b$) and ($c$). A polarization dependence of opposite sign is clearly observed for the two first-order magnetic peaks originating from the striped pattern, this being the classic signature of a chiral magnetic domain morphology.\\
To model the observed scattering behavior, we started from the micromagnetic simulations mentioned before and, using the recipe of Flewett et al. \cite{Flewett19,Flewett17,Loh11}, generated an artificial 2D disordered stripe domain pattern over an 512$\times$512 nm$^2$ surface at 5~nm pixel size. We then used the generalized Fresnel formulae, derived in Ref.~\onlinecite{Qiu2000}, to calculate the reflection (and transmission) coefficient tensor at each point of the surface of the simulated stripe domain sample, for each incident beam angle and magnetization direction. 

Due to limitations of computational power and to the nonlinear scaling of the micromagnetic simulations, these simulations must be performed over an area much smaller than the incident beam size. Diffraction is however determined by the transverse coherence length of the beam, which in our case is about 1~$\mu$m in size\cite{Spezzani2013,Sacchi2007exp}. The algorithm of Flewett et al. \cite{Flewett19,Flewett17,Loh11} used for extending a reduced size micromagnetic simulation over a larger area, assumes that the magnetization vector depends only upon the distance from any given domain wall and on the spatial orientation of such a wall (Bloch, N\`eel or intermediate).

The final step in the simulation process is to calculate the scattering pattern, which is performed simply using Fresnel propagation for a Gaussian beam reflecting from the sample surface according to the already computed array of reflection coefficients. The different polarization steps are treated separately, and finally combined in the calculation of the scattered intensity. In most other XRMS applications, a multilayer sample is used which would require the calculation of reflection coefficients for each and every interface, along with a depth dependent summation considering the attenuation of the beam within the sample. In our case however, with a single layer sample, the only appreciable reflection occurs at the upper surface greatly simplifying the calculation. 

\begin{figure}
 \centering\includegraphics[width=1.0\columnwidth]{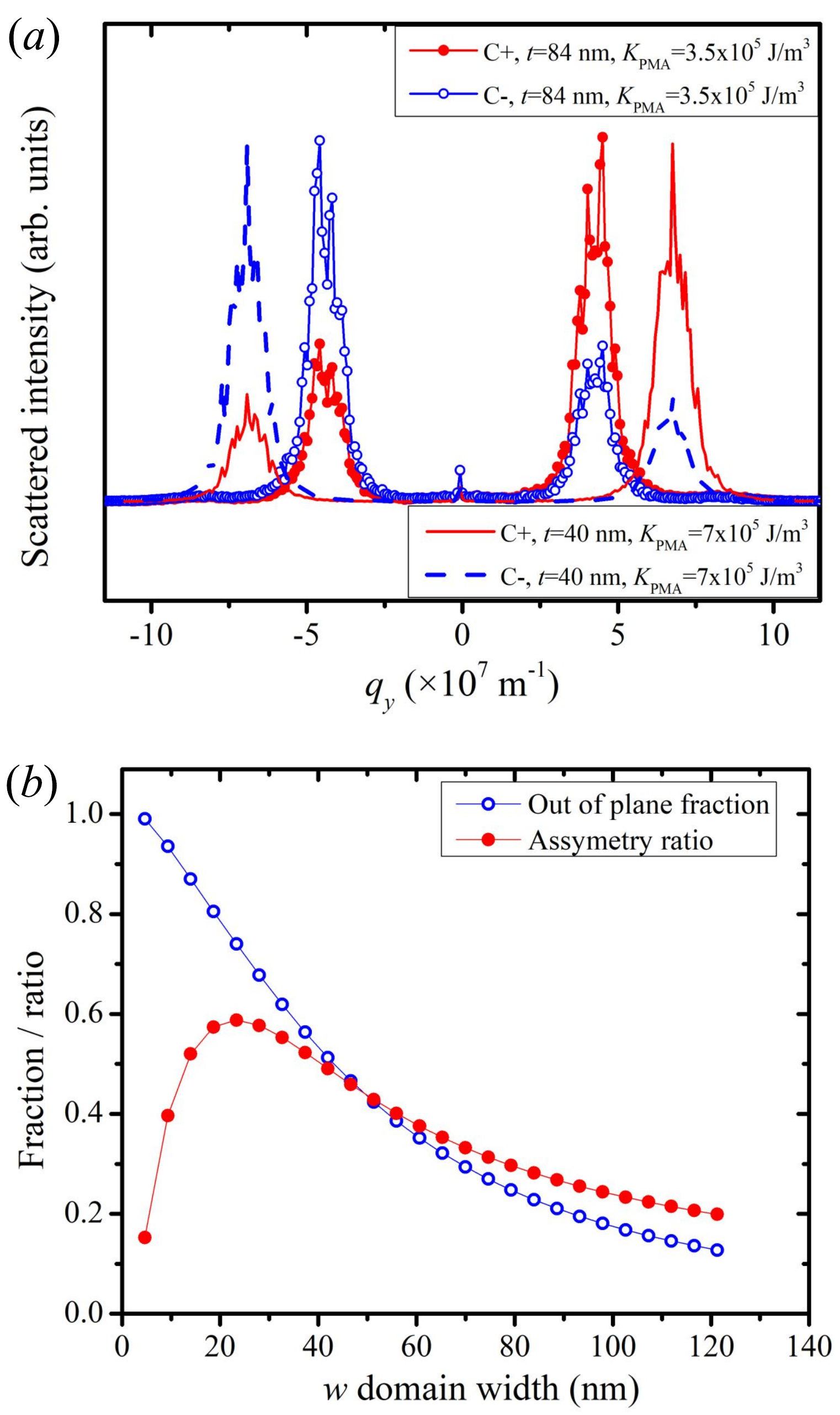}
\caption{\label{fig:Asim} ($a$) Scattered intensity $vs$. $q$ calculated from the magnetization structure calculated by OOMMF. ($b$) Assymetry ratio and out of plane magnetization component as a function of the domain wall width.}
\end{figure}

In Fig.~\ref{fig:Asim}($a$), we show the simulated magnetic diffraction pattern integrated along the direction parallel to the stripes using the micromagnetic simulation with parameters $M_s$~=~1.4$\times$10$^5$~J/m$^3$, $K_{\rm PMA}$~=~3.5$\times$10$^5$~J/m$^3$ and $A$~=~2.9$\times$10$^{-11}$J/m. (solid line). The surface magnetization of this simulation is 30\% out of plane, 15\% in-plane parallel to the incident plane, and 55\% in-plane perpendicular to the incident beam. For comparison, in dashed lines, we show the simulation for a ficticious sample where the parameters were modified with respect to those in agreement with the experimental data. In this case we create a 40~nm thick sample with the same value of $M_s$, and with $K_{\rm PMA}$~=~7.0$\times$10$^5$~J/m$^3$ and $A$~=~2.9$\times$10$^{-11}$~J/m. Due to the higher $K_{\rm PMA}$ value an out of plane component of 59\% was observed, with 37\% in plane perpendicular to the beam, and 4\% in plane parallel to the beam. This value of the in plane/out of plane ratio is reflected in the fact that the asymmetry of the reflection peaks for the 40~nm sample in Fig.~\ref{fig:Asim}($a$) is near to its theoretical maximum for this level of stripe disorder. The simulations performed with the experimental parameters exhibit a lower degree of asymmetry in accordance with its greater in-plane surface component compared with that of the ficticious sample. The aim of the XRMS measurements and simulations was to estimate the out of plane/in plane ratio present, to which effect we additionally simulated the asymmetry ratio as a function of out of plane component using a faster simulation script replacing the micromagnetic simulations with an analytical form of the N\`eel type domain walls given by $M_z$~=~${\rm tanh}$($x$/$w$) where $w$ is the domain wall width and $x$ is the perpendicular distance from the domain wall. $M_x$ was chosen in order to maintain a constant magnitude of the magnetization vector $M$. In Fig.~\ref{fig:Asim}($b$), we show the asymmetry calculated as a function of the domain wall width $w$, using the same degree of disorder as for the experimental case. Here, we observed that the maximum asymmetry ratio is reached for a 20~nm wide domain wall. In the process of performing these simulations we also observed that the asymmetry ratio depended upon the degree of stripe disorder, with lower levels of disorder exhibiting greater levels of asymmetry. To our knowledge, this relation has not been systematically studied in the literature and could warrent investigation. From these experimental results and corresponding simulations we may conclude that there is unequivocal evidence of a chiral N\`eel domain structure present on the surface of the sample, and we estimate that the experimental out of plane component to be between the two cases studied with simulations shown in Fig.~\ref{fig:Asim} - i.e. with between 30\% and 60\% out of plane. The very high uncertainty range stated here is due to the high background from the specular reflection peak which made estimating accurately the asymmetry ratio almost impossible. Readers should note that these figures refer only to the surface layer, and that the out of plane fraction of the bulk is likely to be higher due to the classic triangular shape of the closure domains in samples of this type.

The weak reflection signal in this case can be due to the fact that we were working with a single layer and not a multilayer sample where the reflection signal can be greatly amplified at certain Bragg angles. After the beamtime and shortly before submission of this manuscript, similar samples were characterized using transmission geometry XRMS \cite{Flewett19}, which would have likely been a better candidate than reflection geometry XRMS for samples of this type due to sensitivity to the bulk and not just the surface of the film. 

\subsection{Magnetotransport}
\begin{figure}
 \centering\includegraphics[width=1.0\columnwidth]{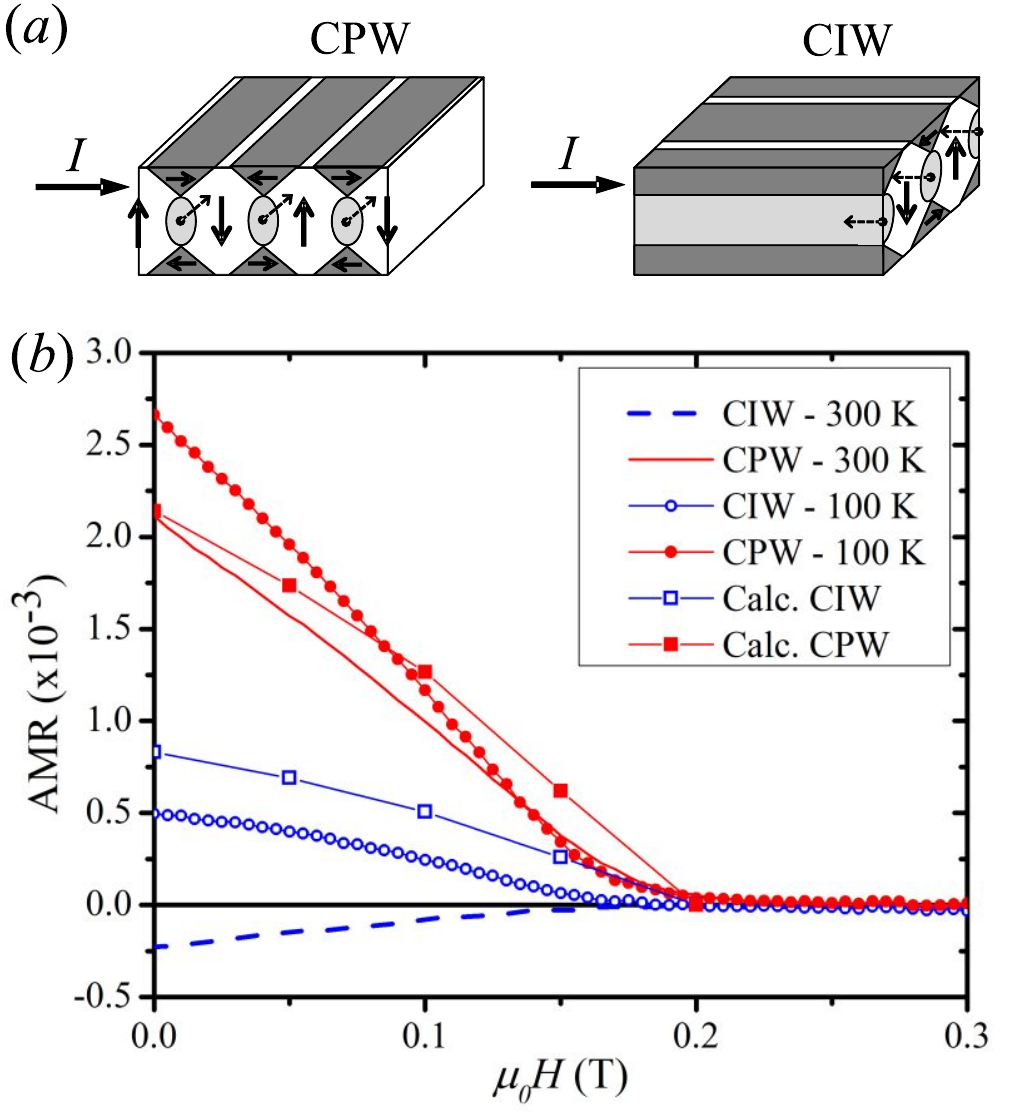}
\caption{\label{fig:AMR}($a$) Geometries used for magnetotransport measurements. ($b$) AMR ratio in both measurement geometries (CPW and CIW) at 300~K and 100~K. The lines wih squared symbols indicate the AMR calculated from simulations.}
\end{figure}

\begin{figure}
 \centering\includegraphics[width=1.0\columnwidth]{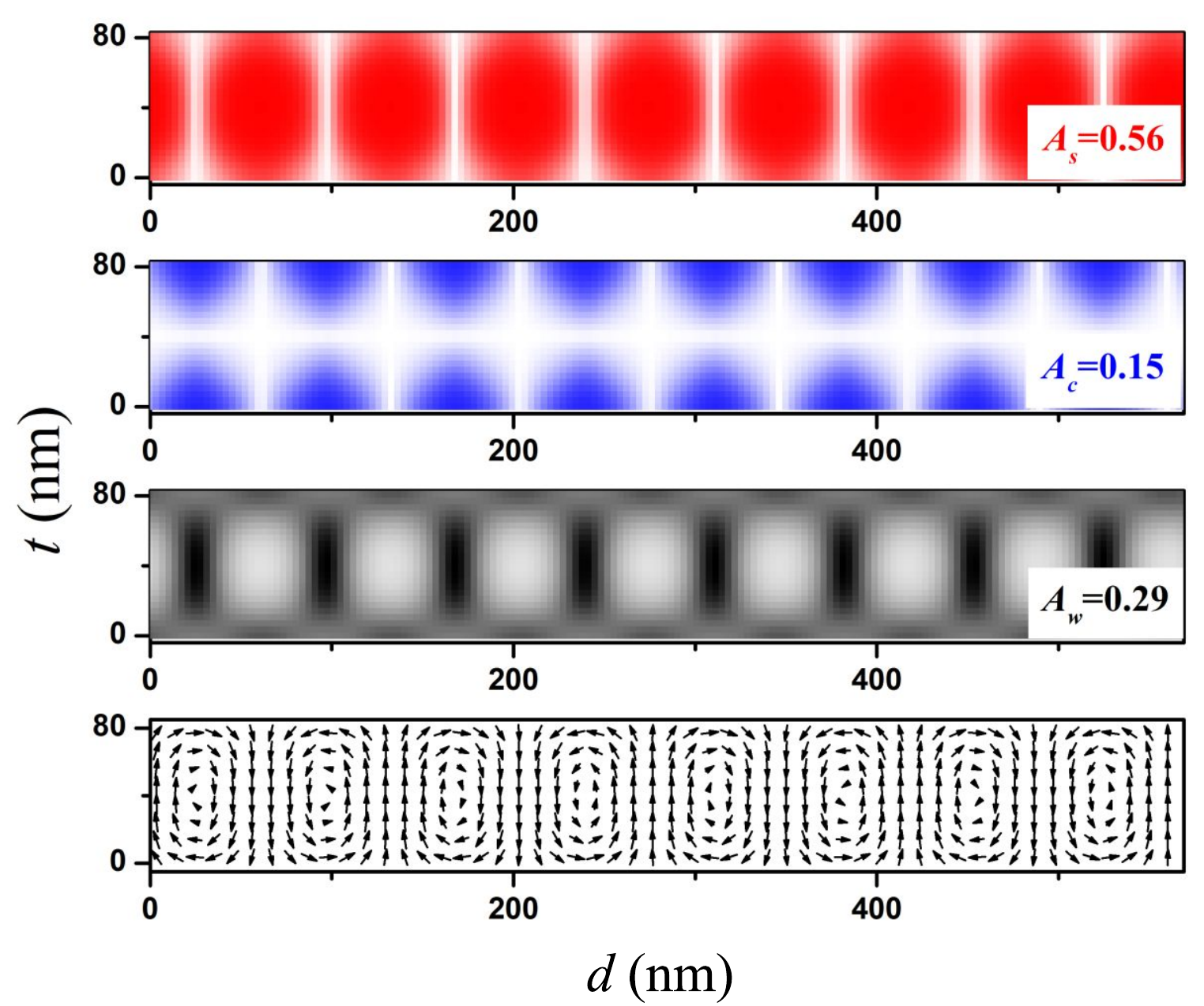}
\caption{\label{fig:VolRat} Cross sectional view of the stripe domains. Three top panles: Color maps for the three components of the magnetization and domain volume ratios calculated by OOMMF package. Bottom panel: Magnetization vector along the sample.}
\end{figure}

In Ref.~\onlinecite{PRB-Mara}, we have studied the magnetotransport properties of this system as a function of temperature for two geometries as depicted in Fig.~\ref{fig:AMR}($a$). In the left panel of that figure we sketch the geometry where the electric current flows perpendicular to the stripe direction (CPW geometry) and in the right panel, we show the geometry where the current flows parallel to the stripe (CIW one). Fig.~\ref{fig:AMR}($b$) diplays the measured MR ratio (defined as MR~=~$\frac{\rho(H)-\rho_{\rm sat}}{\rho_{\rm sat}}$) for the case of CPW and CIW geometries at $T$~=~300~K and 100~K, previously reported in Ref.~\onlinecite{PRB-Mara}. In the case of CPW geometry (red line), we observe that MR-CPW decreases with the applied field irrespective of the temperature.  For the case of CIW, we find a different behavior: at 300~K MR-CIW is positive (dashed blue line), while at 100~K, MR-CIW is negative (open blue circles). In order to explain the sign of MR, some of the authors proposed a simple model\cite{PRB-Mara} based on considering only the AMR contribution of the different kinds of domains as described in Fig.~\ref{fig:1-PreCharac}($a$). Following this model we had success in explaining that MR-CPW can be negative when the electronic current flows perpendicular to the stripes. To understand this behavior, we took into account that the measured resistivity for the three kinds of domains present the following characteristics:\cite{PRB-Mara} $\rho_s > \rho_c > \rho_w$, where subscripts $s$, $c$ and $w$ correspond to those shown in Fig.~\ref{fig:1-PreCharac}($a$). For the perpendicular geometry, at saturation, the resistivity is that of the $w$-type domains ($\rho^{\rm sat}_{\perp}$~=~$\rho_w$) which have the lowest resistivity in comparison with the other domains. It means that the system goes from a higher resistivity state composed of the sum of all the domains (when the stripes are set) to another one with lower resistivity, because at saturation the only domain present is $A_w$, which has the lowest resistivity.\\
On the other hand, when the current flows parallel to the stripes, the presence of the closure domains gives the possibility for the low field MR to be positive or negative. To deal with this feature, we proposed a model of parallel resistors\cite{PRB-Mara} taking into account only the AMR contribution. We have found that if the ratio of the volume fraction of the out of plane domains to the volume fraction corresponding to closure domains, $A_s$/$A_c$, is greater than $\sim$0.6, AMR will be negative, while if such a ratio is lower than $\sim$0.6, AMR will be positive.
We are unable to determine experimentally the domain volume fraction, hence with the aim of estimating it, we used the micromagnetic calculations performed for our samples. To estimate the volume fraction we have considered that the relative volume of each domain is given by  $A_i=M^2_i/M^2$, where $i$ labels domain direction, and $A_s + A_c + A_w = 1$. In Fig.~\ref{fig:VolRat}, we display the calculated volume fraction for the three kinds of domains at remanence.
By taking into account the simulation, the volume ratio of the out of plane domain to the closure one, $A_s$/$A_c$, is $\sim$3.7 which is very far from the volume ratio required for obtaining the AMR inversion (less than $\sim$0.60). Also, we have performed supplementary calculations by slightly changing the magnetic parameters ($M_s$ and $K_{\rm PMA}$ less than 10\%), in order to study the changes in the volume ratios. We have observed that the volume fraction is almost unaffected when such parameters are modified. This would mean that our simple model of resistors is insuficient for explaining the CIW-MR. 

In order to obtain a more realistic picture that allows a better quantitative description of the magnetotransport, from the magnetic structure calculated via OOMMF, we proceed to calculate, via the finite element code COMSOL Multiphysics\cite{Comsol}, the resistivity in the perpendicular and parallel configurations for several values of applied field with the aim of calculating the MR ratio. The modeled sample was the same as the one used in micromagnetic simulation,i.e., a 3-D block of 570$\times$570$\times$84 nm$^3$. The mesh was a uniform tetrahedral elements grid, extruded along the stripe axis with a 4:1 aspect ratio. Approximately 112000 elements and 157000 degrees of freedom were used in each simulation. Volume simulation obeyed the ${\bf J} = \sigma {\bf E}$ law using an isotropic conductivity that was evaluated by interpolating the conductivity map. Boundary conditions were a fixed electric potential difference applied between two opposite edge faces (${\bf J} = \sigma {\bf E}$, Dirichlet condition), and all other boundaries were set to electrical insulation ($\sigma . n$ = 0, Neumann condition). The current density was integrated over one edge face to achieve the total current flow and the electrical resistance.
The results are shown in Fig.~\ref{fig:AMR}(b). From this figure, we can observe a good agreement with the experimental data in the perpendicular direction. While, when the electric current flows parallel to the stripes, the MR-CIW sign obtained from the calculations is opposite to the one observed in the experiments at room temperature, showing that the model fails to predict the MR-CIW. However, for the case of MR-CIW at low temperature (blue dots), the model predicts the correct MR sign (open circles).\\ As it was initially stated, the calculation accounts for the AMR behavior within the stripe phase. Then, from the numerical results, it is possible to study which is the MR fraction that corresponds to the AMR. For the case of CPW at room temperature we observe a good agreement with the calculated AMR values. This indicates that the MR observed in this geometry arises mainly from AMR, and other sources can be considered as negligible. On the other hand, in the case of CIW, the calculated MR is positive, showing the opposite behavior with respect to the room temperature results and indicating the presence of other sources of MR. In the literature, several additional contributions to MR in ferromagnetic films have been studied. In Ref.~\onlinecite{Knittel2005}, the authors show that the Lorentz magnetoresistance (LMR) is negligible when the electronic mean free path, $l_{\rm mfp}$, is much smaller than the domain size. This is our situation because the estimated mean free path at room temperature is $\sim$8~nm and the domain about 30~nm\cite{mariselJPCM}. Also, in that work it is investigated how the internal field of the stripe domains affects the carrier transport. They found that the net result on the carrier is smoothed due to the different directions of the internal field generated by the domains.\\ DW's arise as an important source of MR as well. To understand how the DW's contribute to MR, the different models proposed in the literature deal with how the electronic spin tracks the change of magnetization direction along the wall, and how it is reflected in the resistance \cite{Levy1997,Tatara1997,Reeve2019,Marrows2005}. In our case we have two conditions to fulfill. On one hand, the contribution of DW's to MR is negative at room temperature and, on the other hand, this contribution must change its sign with temperature. R. P. van Gorkom  {\it et  al}. in Ref.~\onlinecite{vanGorkom1999} have developed a model where the DWMR sign depends on the difference of the relaxation times between both spin channels, $\tau^{\uparrow(\downarrow)}$. As it is known, $\tau$ is highly dependent on temperature and this dependence is not the same for both spin channels. This gives the possibility for the MR to change its sign with temperature. Unfortunately, the transport properties in Fe$_{1-x}$Ga$_x$ alloys have not been studied enough in the literature to go into depth in our particular case. Both electronic band structure and impurity scattering contributions must be taken into account to deal with the observed behavior of DWMR.

\section{Conclusions}

In this work, we have demonstrated experimentally by means of XRMS the existence of closure domains in Fe$_{0.8}$Ga$_{0.2}$ thin films where stripes are present, and from these experimental data, with the aid of micromagnetic simulations, have determined that closure domains occupy (55$\pm$15)\% of the surface magnetization. 
The micromagnetic calculations indicate the model of volume fraction proposed in Ref.~\onlinecite{PRB-Mara} is suitable when the electronic current flows perpendicular to the stripes, MR-CPW, suggesting that the main contribution to MR-CPW is from AMR. For current flowing parallel to the stripes, MR-CIW, such a model fails to explain the sign of MR at room temperature, while at low temperature, the model predicts the experimental observed behavior. This shows that an intrinsic contribution to the MR from domain walls must be taken into account in order to explain the positive MR observed.
Further detailed magnetotransport experiments as a function of temperature, band structure calculation and also modeling of scattering rates are needed in order to go into depth in terms to understanding MR in Fe$_{1-x}$Ga$_x$ alloys.

\begin{acknowledgments}
The authors gratefully acknowledge S. Bustingorry for fruitful discussions about this work and also R. Benavides, C. P\'erez and M.~Guill\'en for technical support. B.~P., M.~B. and J.~M. acknowledge partial financial support by PIP 11220120100250CO, PICT 2013-0401 and SIIP 06/C510.

\end{acknowledgments}

\begin{thebibliography}{43}
\bibitem [{hag()}]{hagostripes}%
  \BibitemOpen
  \href@noop {} {}\bibinfo {note} {By applying a magnetic field in the sample
  plane and then releasing it, stripe domains align along the direction of the
  applied field}\BibitemShut {NoStop}%
\bibitem [{exp()}]{expSR}%
  \BibitemOpen
  \href@noop {} {}\bibinfo {note} {The specular spot is enlarged by the
  presence of weak non-monochromatic low-energy photons (the fundamental energy
  of the wiggler source is about 5~eV) that are reflected with high efficiency
  from the sample.}\BibitemShut {Stop}%
\end{thebibliography}
%
\end{document}